# CELL ARCHITECTURE FOR NANOELECTRONIC DESIGN

*Ferran Martorell and Antonio Rubio*

High Performance IC Design Group, DEE, U. Politècnica de Catalunya
Ed. C-4 Jordi Girona 1-3 08034, Barcelona, Spain
Corresponding author e-mail: ferranm@eel.upc.edu

## ABSTRACT

Several nanoelectronic devices have been already proved. However, no architecture which makes use of them provides a feasible opportunity to build medium/large systems. Nanoarchitecture proposals only solve a small part of the problems needed to achieve a real design. In this paper, we propose and analyze a cell architecture that overcomes most of those at the gate level. Using the cell structure we build 2 and 3-input NAND gates showing their error probabilities. Finally, we outline a method to further improve the structure's tolerance by taking advantage of interferences among nanodevices. Using this improvement we show that it is possible to reduce the output standard deviation by a factor larger than $\sqrt{N}$ and restitute the signal levels using nanodevices.

## 1. INTRODUCTION

Nanotechnology environment is expected to be very different than today CMOS design space. Nowadays technology assumptions of defect free designs and zero error determine the design methodology. However, in the nanoscale these assumptions fail. The reduction of minimum dimensions makes difficult the fabrication process resulting in high defect ratios and large variations on device parameters. Besides, integration density increase forces a reduction of signal levels reducing the noise margins of the technology. This fact results in device internal noise limiting the system performance. Furthermore, thermal noise, flicker noise, cross-couplings and ground noise need to be considered in order to design a reliable nanoarchitecture. On top of all, the reduction of the device dimensions also increases the probability of error due to particle interactions with groups of nanodevices. Reliable nanoarchitectures should consider all these problems in a structure practical for implementation.

Several architectures at different system levels have been proposed to build systems based on nanoscale devices [1]-[6]. However these architectures only consider some of the problems and obviate all others.

Most works proposing architectures for nanotechnologies consider unreliable devices. Some of them also treat the problem of transient faults due to noise. However, none of them propose a feasible solution for all the problems in the same architecture. The two most relevant architectures proposed which attack the problem from two different points of view are the reconfigurable architectures based in PGA structures and the fault and defect tolerance properties derived from variations of NAND Multiplexing technique proposed by Von Neumman [1]. The former requires testing and configuring all the devices in the design, a very slow and costly process. Besides, no protection against transient faults is implemented. The latter is very attractive by its combined defect and fault tolerance, but it has implementation difficulties (at the current state of the art) as it requires the implementation of several nanoscale interconnections per circuit.

In this work we propose a cell architecture intended to overcome most of them in a structure practical for fabrication (section 2). Section 3 analyzes its defect and fault tolerance properties according to area and energy cost. Section 4 describes the implementation of Boolean gates using the cell architecture and analyze the error probability for 2 and 3-input NAND gates. Section 5 outlines a modification on the cell working principles that may lead to tolerance capabilities better than those obtained by simple averaging and section 6 indicates how to restitute signal using the proposed structure.

## 2. CELL ARCHITECTURE FOR NANOELECTRONICS

We define our cell architecture using similar ideas to [7], but minimizing the interconnection problem. Figure 1 shows the scheme for our cell architecture. It is composed by two units. The first unit is composed by nanodevices that perform the computation operation with loss of information due to noise and defects. This unit is characterized by a high degree of redundancy, $N$. The nanodevices receive a single input, $x_i$, to keep down the complexity of assembling them and reducing the probability of errors in the fabrication process. The second unit restores the signal levels for the computation results, $y_o$. This block produces a restored output, $y_{out}$ and may be built either with MOS technology or nanodevices depending on the system requirements. Cell interconnections are built by metal lines as in CMOS technology.

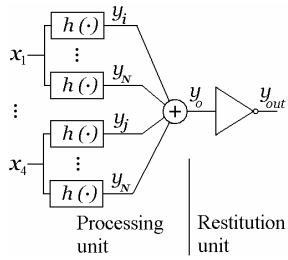

**Fig. 1** Scheme for the proposed cell architecture. The structure is composed by two differentiate units. The processing unit composed by nanodevices and the restitution unit built with MOS technology.

The processing is done following a distributed redundant scheme of threshold logic gate in which each input is introduced in the final adder by a bundle of $Ni$ nanodevices – threshold logic gates have a higher tolerance to faults than Boolean gates [8]. In common threshold gates the cell sums all its inputs scaled by their weights and afterwards a selective threshold is applied to perform a part of processing with restitution of levels. In the proposed cell, the threshold is applied using one fixed input (see section 4 for details). Input weighting is done by adjusting the relative number of nanodevices in each input bundle ($N$=min($N_i$)). The restitution unit is identical for all gates simplifying the design. The cell transfer function has a generic expression of the form

$$y_{out} = \mathrm{sgn}\{\omega_1(x_1 - T_1) + ... + \omega_j(x_j - T_j)\}, \qquad (1)$$

where $\omega_i = N_i/N$ are the weight for each input in the cell, $x_i$ are the cell inputs and $T_i$ are the mean threshold value for each set of nanodevices. This structure is suitable for implementing threshold logic gates and neural computations.

The nanoscale unit of the cell is composed by sets of identical nanodevices in a parallel configuration. To simplify the fabrication process the nanodevices should have common terminals. The fabrication of these structures may be done by a similar method as nanopores (which are structures already fabricated for molecular conductance measurements [9]). These structures are nanoscale holes (diameter of 30 to 50 nm) in which molecules are self-assembled. Assuming a diameter of 50 nm for a nanopore it is possible to build up 90 bundles of about 2500 nanodevices on a single layer filling the area of a 65 nm CMOS inverter (0.19 μm²). As these devices do not rely on the crystalline structure of the silicon several layers may be implemented. The additive function may be implemented by summing up

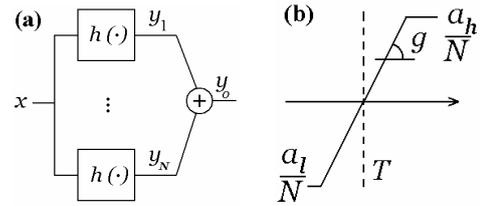

**Fig. 2** (a) Diagram for a single input gate used for analyzing the cell tolerance properties. (b) Generic transfer function modeling the nanodevice response. Two separate states with a linear transition between them.

the charges flowing through the nanodevices on the gate capacitance of the MOS inverter. The restitution unit and the communication stage may be initially built using MOS technology. The area reduction ratio for a given gate may range from 2 to 10 (depending on the circuit complexity). With nanotechnology advances, the gate complexity may be increased further improving the integration density. Eventually, by either improving the fabrication techniques or improving the cell architecture the restitution unit may be built with nanoscale devices increasing the functional integration density by about 2 orders of magnitude.

## 3. TOLERANCE ANALYSIS

In this section we analyze the tolerance capabilities of the proposed cell. The cell architecture is composed by several bundles of nanodevices with a single input as depicted in figure 2(a). In them, an input, $x$, is applied to each nanodevice which performs a function, $h(\cdot)$, producing a partial output, $y_i$. Each device produces a $1/N$ part of the total output function, $H(\cdot)$. Input-output functions for the nanodevices, $h(\cdot)$, are modeled by a simple generic transfer function which considers two stable states and a linear transition between them (figure 2(b)). It reads

$$y_i = h(x) = \begin{cases} a_h/N & x + \eta - T > b_h \\ g(x + \eta - T) & b_h \geq x + \eta - T \geq b_l \\ a_l/N & x + \eta - T < b_l, \end{cases} \qquad (2)$$

where $b_h = a_h/(Ng)$ and $b_l = a_l/(Ng)$. The function is able to model any two state system with a transition region. The two output states are $a_h/N$ and $a_l/N$ (which for simplicity we assume symmetrical), $T$ stands for the threshold or constant offset, $g$ for the equivalent gain of the transition section, $\eta$ for the internal noise and $N$ for the redundancy factor. This structure is representative of the cell because a complete processing cell is simply the linear sum of several intermediate outputs. Therefore, by analyzing a single intermediate output, the cell behavior may be characterized. As our interest is the reliability of the nanoscale section we only analyze its output, $y_o$, and assume that the MOS inverter has a precise high gain threshold.

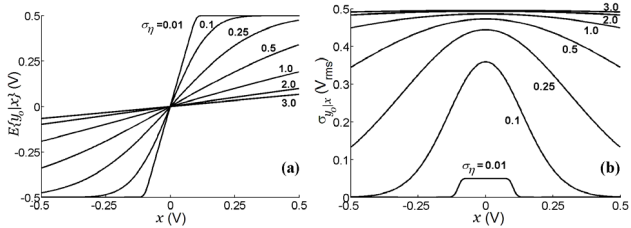

**Fig. 3** Output characteristics for a single element with white Gaussian internal noise with standard deviation $\sigma_\eta$. (a) Expected value and (b) standard deviation.

To perform the tolerance analysis we calculate the expected value and the variance due to each random variable ($\eta$, $T$ and $g$). We proceed by first working out the statistics of the individual outputs, $y_i$, which read

$$E\{y_i|x\} = \int_u h(x) f_u(u) du \qquad (3)$$

and

$$\sigma^2_{y_i|x} = \int_u h^2(x) f_u(u) du - E^2\{y_i|x\}, \qquad (4)$$

where $u$ stands for the random parameter considered and $f_u(u)$ for its probability density function. Then, by considering identically independent distributions among the nanodevices we compute the bundle output statistics by adding up the individual values. All the expressions are conditioned to the input value, $x$.

### 3.1. Noise Error

To analyze the internal noise tolerance of this structure we consider an independent additive Gaussian noise -- with zero mean and standard deviation $\sigma_\eta$ – added to each nanodevice. We characterize the cell by computing the expected value, which provides an estimation of the transfer function, and its output variance, that gives a measure of the noise at the output. A first order simplification of the expressions is enough to appreciate the cell behavior (for a full derivation see [10]). The input-output function is approximated as

$$E\{y_o|x\} \approx \frac{a_h - a_l}{\sigma_\eta \sqrt{2\pi}}(x - T) \qquad (5)$$

bounded between $[a_h, a_l]$ and the input conditioned variance as

$$\sigma^2_{y_o|x} \approx \frac{1}{N}\left[\frac{(a_h - a_l)^2}{4} - \frac{1}{6}\frac{(a_h - a_l)^3}{g\sigma_\eta\sqrt{2\pi}} - \left(\frac{a_h - a_l}{\sigma_\eta\sqrt{2\pi}}\right)^2(x - T)^2\right] \quad (6)$$

Figure 3 plots the full expressions for the expected value and standard deviation (std) for a single element with $T=0$ V, $g=50$ and $a_h=-a_l=0.5$ V. When considering a bundle of elements std curves are divided by a factor $\sqrt{N}$. From (5) and figure 3(a) we observe that the cell produces an output signal which is a non-linear copy of the input signal with a certain gain and shift according to the nanodevice gain and threshold. Noise amplitude reduces the gain and the output levels. This determines the minimum amount of energy required per operation and the minimum working levels for signals. From (6) we observe the effect of redundant circuits. The variance is reduced at least by a factor $N$ (due to the average function). We can see that the variance is not constant (figure 3(b)) for all the input range. Thus, appropriately selecting the working levels a further reduction may be achieved.

### 3.2. Parameter Variation Tolerance

As technology becomes more unreliable, the gradation between working and defective devices spreads producing a wide range of parameter values. In this section, we are interested in the middle range variations. Therefore, we model the random variable by a uniform distribution. Values with a large deviation from the mean value produce non-functional devices that may be considered in section 3.3.

Our simple transfer function, $h(\cdot)$, is characterized by two parameters (threshold and gain) that summarize the physical variations in nanodevices. We assume these parameters to be independent random variables.

#### 3.2.1. Threshold Error
We model the threshold fluctuations with a uniform distribution with mean $\mu_T$ and amplitude $A_T$. As in the previous section, we compute the mean value which reads

$$E\{y_o|x\} = \frac{a_h - a_l}{A_T}(x - \mu_T) \qquad (7)$$

bounded between $[a_h, a_l]$ and the variance

$$\sigma^2_{y_o|x} \approx \frac{1}{N}\left[\frac{(a_h - a_l)^2}{2}\left(1 + \frac{a_h - a_l}{gA_T}\right) - \frac{(a_h - a_l)^3}{12gA_T} - \frac{(a_h - a_l)^2}{A_T^2}(x - \mu_T)^2\right] \qquad (8)$$

The effects of threshold variations are quite similar to noise effects. They produce a constant reduction of the output levels and the variance is at least reduced according to the redundancy factor. Limiting the threshold variations is crucial to produce a functional cell. If fluctuations are too large the cell is not able to differentiate its output states due to the smoothing effect. This problem can not be corrected by increasing the redundancy factor or applying any other known technique.

#### 3.2.2. Gain Error
The gain error only affects the transition section of the transfer function. If signal levels are out of this section its effect is negligible while the gain is sufficiently high. Extremely low gain nanodevices produce nonfunctional elements that can be modeled as defective. The expected value in the linear section reads

$$E\{y_o|x\} = \mu_g(x - T) \qquad (9)$$

bounded between $[a_h, a_l]$ and the variance

$$\sigma_{y_o|x}^2 \approx \frac{1}{N}\frac{A_g^2}{12}(x - T)^2 \qquad (10)$$

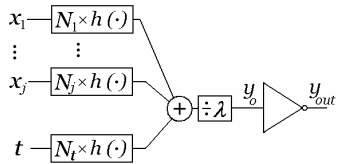

**Fig. 4** Generic logic gate structure using the proposed cell architecture.

From these expressions we observe that the redundancy factor reduces the effects of the gain error and that the mean value is not affected by gain fluctuations. In any case the effects produced by the gain error are small enough to be nearly irrelevant in front of threshold variations.

### 3.3. Defect Tolerance

To analyze the defect tolerance of this structure we consider the main fabrication errors: (*i*) connection errors at the input or output of the nanodevice giving no signal output; (*ii*) malfunctions produced by an extremely low gain, or (*iii*) shorted devices, both modeled as a low conductivity connection between input and output ($y_i=x/N$); and (*iv*) exceedingly high or low threshold values that produce a constant one or zero state on the nanodevice ($y_i=a_h/N$ or $y_i=a_l/N$). The probabilities for each defect furnish a mean number of functional devices *j*, shorted *k*, unconnected *l*, stuck-at-one *m*, and stuck-at-zero *n*. The partial output is then

$$y_o = \frac{1}{N}\left(\sum_j H(x) + \sum_k x + \sum_l 0 + \sum_m a_h + \sum_n a_l\right) =$$
$$= \frac{j}{N}H(x) + \frac{k}{N}x + \frac{m-n}{N}a_h \qquad (11)$$

where $j+k+l+m+n=N$ or in general the number of devices in the bundle. From this equation we observe that defects are converted in a graceful degradation of the cell response. The output levels are reduced and a small offset is added. However, by increasing the number of devices it is possible to partially compensate this problem.

### 3.4. Discussion

The analysis shows the tolerance of the cell to internal noise, parameter variations and defects as a function of the area cost (redundancy factor, *N*) and the energy consumption (output levels, $a_h$ and $a_l$). However, not all problems have been accounted for. Highly correlated noise among the nanodevices – such as cross-coupled interferences or ground bounce noise – or particle interactions – which may affect a great number of nanodevices – are phenomena that require a different approach. Part of these interferences may be eliminated by the inherent noise margin of the cell (see figure 3(a)). It must be noted that the noise margin depends on $\eta$, $T$ and $g$. If the noise margin is not enough, we may use error detection/correction codifications combined with this architecture. It is very difficult to deal with all the problems arising in nanotechnology with a single level of circuits or tolerance strategy. As the origin of those problems is so different, it seems that tolerance mechanisms at several levels may be a better solution [11].

### 4. IMPLEMENTATION OF NAND GATES

After characterizing the cell's tolerance, this section presents the implementation of logic gates using the proposed cell architecture. The logic gates are based on a threshold logic scheme both because of its increased fault tolerance [8] and because the cell structure naturally fits in the threshold logic structure. Figure 4 depicts the general scheme for a j-input logic gate. The inputs, $x_i$, pass through bundles composed by $N_i$ nanodevices producing intermediate outputs that are averaged into $y_o$. A special fixed input, $t$, (high or low) is used to define a variable cell offset that keeps the decision threshold constant at 0 V (the selected decision threshold of the restitution unit). The averaging factor, $\lambda$, is used to keep the middle output values inside the working region (between $a_h$ and $a_l$). In these examples, we consider the restitution unit built from a MOS inverter. Then, $y_o$ must produce an inverted logical function in order to obtain the desired Boolean gate. The number of nanodevices in each bundle, $N_i$, is a multiple of the main redundancy factor, $N$, and are used to produce the necessary weights among the gate inputs according to threshold logic design style [12]. Therefore, the nanodevices are used to adjust the weight of each input.

As an example two NAND gates with 2 and 3 inputs are designed and simulated. As NAND logic function is able to generate any complex binary function it also proves the universality of this cell architecture. Due to the MOS inverter, the middle output, $y_o$, must implement an AND function. For implementing a 2-input NAND the cell parameters are $N_1=N_2=N_t=N$, $\lambda=3$ and t=a*l*. For a 3-input NAND they are $N_1=N_2=N_3=N$, $N_t=2N$, $\lambda=5$ and $t=a_l$. NOR gates may be obtained by simply setting $t=a_h$. We simulate the response of these gates only considering the effects of internal noise ($\sigma_\eta$) as it represents the effects of noise and threshold variations (their effects are additive). In this analysis we obviate the effects of defects that will increase the error probability by reducing the output levels.

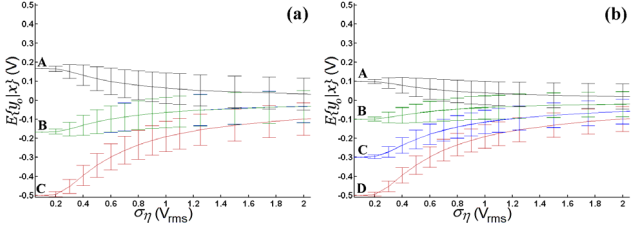

**Fig. 5** Output at a probability of 99.7% for a redundancy factor $N$=100. (a) Two-input AND function (A=11, B=01,10 and C=00). (b) Three-input AND function (A=111, B=011,101,110, C=001,010,100 and D=000).

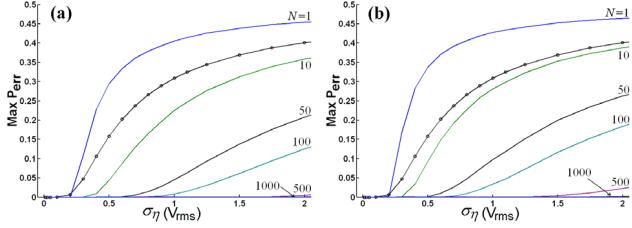

**Fig. 6** Maximum error probability (worst case input combination) for the NAND gate for several redundancy factors ($N$=1, 10, 50, 100, 500 and 1000). (a) Two-input AND gate. (b) Three-input AND gate.

Figure 5 shows the mean output (continuous lines) for each input combination against $\sigma_\eta$ for gates with $N$=100. The error bars cover 99.7% of the outcomes ($\pm 3\sigma_{yo}$). Plot (a) corresponds to the 2-input gate (where input combinations are A=11, B=01,10 and C=00) and (b) to the 3-input gate (A=111, B=011,101,110, C=001,010,100 and D=000). Feeding these signals into the MOS inverter we obtain a NAND gate able to produce a correct output for a large range of internal noise amplitudes. Figure 6 presents the actual error probability for those ((a) 2-input and (b) 3-input). The figure depicts the error probability for several redundancy factors ($N$=1, 10, 50, 100, 500 and 1000). To provide a reference the -o- line shows the probability of a Gaussian distributed signal with mean $a_h$ and std $\sigma_\eta$ to cross the 0 V threshold. These results along with the previous analysis indicate that the cell architecture is able to successfully work in the nanoscale region.

## 5. ENHANCED NAND GATES

NAND gates presented in section 4 are able to work for a wide range of noise amplitudes by taking advantage of the standard deviation reduction factor of $\sqrt{N}$. However, the smoothing effects of noise and threshold quickly reduce the output amplitudes degrading the gate performance and limiting the maximum function

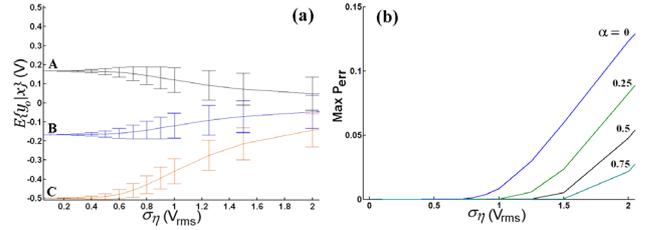

**Fig. 7** (a) Output at a probability of 99.7% for a two-input AND function with $\alpha$=0.5 and $N$=100. (b) Error probability for an enhanced 2-input AND gate ($N$=100 and $\alpha$=0.0, 0.25, 0.5, 0.75).

complexity in the processing unit. In this section, we describe how it is possible to reduce noise smoothing and achieve a std reduction factor larger than $\sqrt{N}$ by taking advantage of interferences among nanodevices. Similar results have been also reported in neurons [13].

Due to the proximity of nanodevices, the actual state of one of them may be affected by their close neighbors. Similarly to magnetization effects in the Ising model. We have computed this interaction by considering that the actual nanodevice threshold $T_i$ depends on the states of the 4 nearest nanodevices according to

$$T_i = T + (a_h - a_l) \left( \sum_{low} \frac{\alpha}{2} - \sum_{high} \frac{\alpha}{2} \right) \quad (12)$$

Thus, the actual threshold is the sum of the device static threshold, $T$, and a fraction of the working range, $a_h$-$a_l$. This fraction is adjusted by the interaction strength, $\alpha$, from $2\alpha$ (all neighbors at low state) to -$2\alpha$ (all high). Considering this local interaction, the cell performance is greatly improved. Figure 7(a) shows the mean output levels of a 2-input NAND gate with 10x10 nanodevices and $\alpha$=0.5. Comparing figure 5(a) to 7(a) the reduction of noise smoothing and a std reduction factor larger than $\sqrt{N}$ are clear. Figure 7(b) shows how the error probability for a bundle of 10x10 nanodevices reduces with an increasing interaction strength ($\alpha$=0.0, 0.25, 0.5 and 0.75).

## 6. NANOSCALE RESTITUTION UNITS

To be able to build restitution units using nanoscale devices a structure able to produce a restituted and reliable output (i.e. a high gain cell with a low output error probability) is necessary. Due to low gain one cell cannot provide the required gain to restitute an output. To solve this limitation we propose to use chains of cells in order to achieve the necessary gain. We have simulated the response of chains with 5 gates with interaction strength of $\alpha$=0.5 for different noise levels. The results are presented in fig. 8 where we observe the mean output of the first and last gate along with the probability of degrading the signal (i.e. $|y_o|<|x|$) at the fifth gate (left to right plots). The cells without interaction are only able to produce reliable outputs up to $\sigma_\eta$=0.2 V$_{rms}$ while cells with $\alpha$=0.5 are able to define two stable and reliable output states up to $\sigma_\eta$=0.8 $V_{rms}$. These results indicate that it is possible to reduce the gate dimensions by implementing the restitution unit using chains of several cells. The number of cells required will depend on the requirements of each gate.

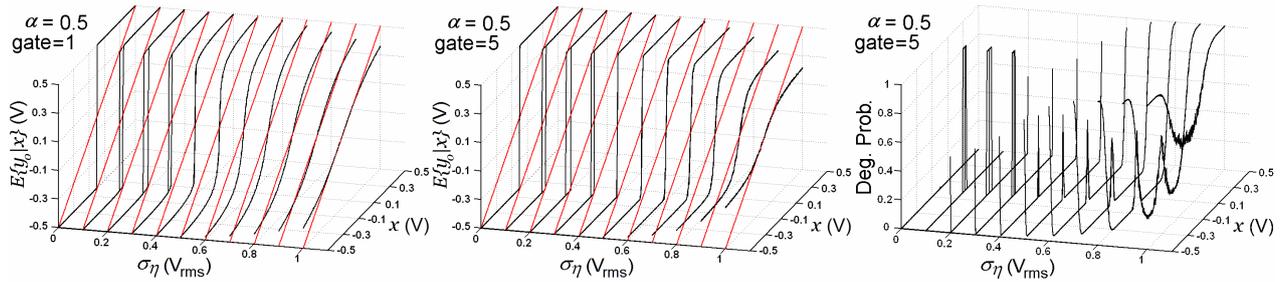

**Fig. 8** Response of chains of 5 cells with interaction $\alpha$=0.5. From left to right, expected output at the first cell, expected output at the fifth cell and degradation probability ($|y_o|<|x|$) at the fifth gate.

## 7. CONCLUSION

The main problems that nanotechnologies need to address are identified. Considering all aspects of the complex design space appearing with the nanoscale, an alternative architecture at the cell level is presented and its defect and fault tolerance are derived as a function of the area and energy cost. Special care is put in the fabrication feasibility of this structure resulting in a combination of nanodevices structured in nanopores and when necessary MOS inverters. The architecture tolerance properties are shown by designing and simulating two NAND gates with 2 and 3 inputs. Their error probabilities are calculated as a function of the area cost ($N$) showing a wide working range. Furthermore, we outline a method to further improve these results by taking advantage of the interferences among close nanodevices. This method reduces the loss of information due to noise smoothing and provides a std reduction factor larger than $\sqrt{N}$. Finally, by considering chains of cells with interaction it is possible to build fully nanoscale gates which permit a quantitative improvement on the miniaturization of this cell architecture. All these results suggest that this cell architecture may be valid for the nanoscale.

## 8. ACKNOWLEDGEMENT

This work is funded by the grant AP2002-2600 and the project TEC2005-02739/MIC.